\documentclass[sigconf]{acmart}

\usepackage{multirow} 
\usepackage{xspace} 
\usepackage{xcolor} 
\usepackage{graphicx} 
\usepackage{subcaption} 

\usepackage{xcolor,colortbl}
\usepackage{enumitem} 

\usepackage{dirtytalk} 

\usepackage{booktabs}
\definecolor{rowgray}{HTML}{F1F5F9}
\newcommand{\sh}{\cellcolor{rowgray}}

\newcolumntype{L}[1]{>{\raggedright\arraybackslash\hyphenpenalty=10000\exhyphenpenalty=10000}p{#1}}
\newcolumntype{M}[1]{>{\raggedright\arraybackslash\hyphenpenalty=10000\exhyphenpenalty=10000}m{#1}}

\usepackage{tikz}
\usepackage{array}
\usepackage{stfloats}

\newcommand{\sayit}[1]{{\textit{\say{#1}}}}

  {\list{}{\selectfont\itshape\leftmargin=0.25in\rightmargin=0.25in}\item[]}%
  {\endlist}

\AtBeginDocument{%
  \providecommand\BibTeX{{%
    \normalfont B\kern-0.5em{\scshape i\kern-0.25em b}\kern-0.8em\TeX}}}

\newcommand{\SystemName}{\textsc{MaterialVellum}\xspace}

\usepackage{caption}
\copyrightyear{2026}
\acmYear{2026}
\setcopyright{cc}
\setcctype{by}
\acmConference[UIST '26]{The 39th Annual ACM Symposium on User Interface Software and Technology}{November 02--05, 2026}{Detroit, MI, USA}
\acmBooktitle{The 39th Annual ACM Symposium on User Interface Software and Technology (UIST '26), November 02--05, 2026, Detroit, MI, USA}
\acmDOI{10.1145/3830398.3830504}
\acmISBN{979-8-4007-2856-3/2026/11}

\begin{document}

\sloppy

\title[MaterialVellum: Steering Generative AI by Leveraging Expressive Material Affordances
to Form Design Constraints]{MaterialVellum: Steering Generative AI by Leveraging Expressive Material Affordances
to Form Design Constraints}

\settopmatter{authorsperrow=4}

\author{Chu Li}
\affiliation{
  \institution{Microsoft Research}
  \city{Redmond}
  \state{WA}
  \country{USA} 
}
\affiliation{
  \institution{University of Washington}
  \city{Seattle}
  \state{WA}
  \country{USA} 
}
\email{chuchuli@cs.washington.edu}

\author{Nicolai Marquardt}
\affiliation{
  \institution{Microsoft Research}
  \city{Redmond}
  \state{WA}
  \country{USA} 
}
\email{nicmarquardt@microsoft.com}

\author{Hugo Romat}
\affiliation{
  \institution{Microsoft Research}
  \city{Redmond}
  \state{WA}
  \country{USA} 
}
\email{romathugo@microsoft.com}

\author{Ken Hinckley}
\affiliation{
  \institution{Microsoft Research}
  \city{Redmond}
  \state{WA}
  \country{USA} 
}
\email{kenh@microsoft.com}

\author{Nathalie Riche}
\affiliation{
  \institution{Microsoft Research}
  \city{Redmond}
  \state{WA}
  \country{USA} 
}
\email{nath@microsoft.com}

\author{Kwon Ko}
\affiliation{
  \institution{Microsoft Research}
  \city{Redmond}
  \state{WA}
  \country{USA} 
}
\affiliation{
  \institution{Stanford University}
  \city{Stanford}
  \state{CA}
  \country{USA} 
}
\email{kwonko@stanford.edu}

\author{Fanny Chevalier}
\affiliation{
  \institution{Microsoft Research}
  \city{Redmond}
  \state{WA}
  \country{USA} 
}
\affiliation{
  \institution{University of Toronto}
  \city{Toronto}
  \state{ON}
  \country{Canada} 
}
\email{fanny@dgp.toronto.edu}

\renewcommand{\shortauthors}{Li et al.}


\begin{abstract}

We present MaterialVellum, an interaction approach for steering generative AI through material-inspired manipulation of design constraints.
In urban design, for instance, zoning bounds can be defined by the stretch of elastic bands, building density redistributed by molding clay, and functional relationships organized by the pull of magnets.
MaterialVellum departs from current interfaces that rely on natural language prompts, and instead leverages users' existing knowledge of physical materials to represent and manipulate design constraints. 
We introduce a framework that maps material affordances---such as elasticity, malleability, or attraction---to spatial, relational, and proportional constraints that underpin design. 
These mappings are operationalized in user interfaces through familiar direct manipulation multi-touch gestures, including stretching, sculpting, and clustering, whose effects feed as control parameters into generative AI pipelines. 
We demonstrate this approach through technology probes, including an urban design, a flower arrangement design, and a graphical design scenario.  
A qualitative study (N=10) shows that material-based interactions improve accessibility of constraint expression, support exploratory workflows, and lower barriers for non-experts, while effectiveness depends on alignment with users' mental models.

\end{abstract}



\begin{teaserfigure}
  \centering
  \includegraphics[width=\textwidth]{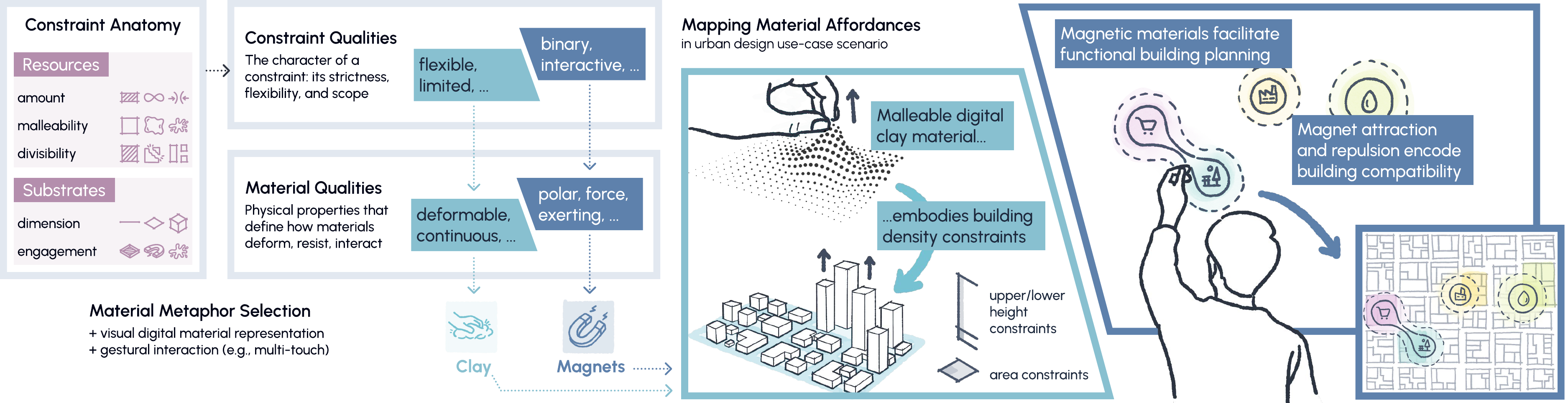}
  \vspace{-8pt} 
  \caption{\SystemName: (left) Framework mapping material affordances to spatial and relational constraints in design, (right) which is demonstrated in urban design and other use case scenarios.}
  \vspace{5pt}
  \Description{Left, the framework: a Constraint Anatomy panel listing resources (amount, malleability, divisibility) and substrates (dimension, engagement) feeds into paired panels of constraint qualities and material qualities, which converge on material metaphor selection, illustrated with clay and magnet icons. Right, the same mapping in urban design: hands pinch a dot grid of digital clay encoding building density, and drag colored magnets whose attraction and repulsion encode building compatibility, producing a generated urban plan.}
  \label{fig:teaser}
\end{teaserfigure}

\maketitle


\section{Introduction}

Design constraints are central to the creative process. They define the creators' arena by specifying both the boundaries of the solution space and identifying the open spaces where artistic interpretation can flourish~\cite{onarheim_opening_2010}. 
Such constraints---hard or soft---take many forms: spatial grids that organize web interfaces, height regulations that shape urban skylines, and proportional rules that define fashion design. 
Creative professionals then shape solutions within this bounded space and iterate over solutions, balancing tradeoffs against given constraints.

When creators co-design with generative AI (genAI), design constraints further function as an expressive way for communicating the guardrails for steering a model's creative expression. 
Still, while natural language is effective for many expressions of intent, the articulation of constraints---from finding the right terminology to express, to ways of setting their scope---can be tedious~\cite{subramonyam_bridging_2024}. 
This is especially true of non-textual media where sketching (e.g., a grid line), modeling (e.g., a building's shape), fitting (e.g., pieces of fabric), might come as more direct and natural means of expression. In this work, we explore the potential of leveraging physical material affordances to shape design constraints within co-design workflows with genAI, as an alternative to the classic textual prompts.

Reasoning about constraints (e.g., flexible-rigid, soft-hard) aligns conceptually with reasoning about material behavior (e.g., elastic-solid, porous-dense). We exploit this alignment by proposing a conceptual mapping that links material qualities---such as elasticity, density, attraction---to the direct expression of constraints. For instance, consider an urban designer planning for \textit{20 to 25 acres} of green space. This constraint can be conceptualized as a rubber band: its ideal tension represents the target area, while slackness or snapping indicates a failure to meet that value. Here, the physical properties of the rubber band embody the constraint's tolerance zone. 
We conceptualize constraints as interactions between \textit{resources} (the elements being constrained) and a \textit{substrate} (the space within which constraints operate). Resources are categorized by their amount, malleability, and divisibility, whereas substrates are defined by their dimensionality and their active or passive nature. These mappings are operationalized in user interfaces through familiar direct-manipulation multi-touch gestures, including stretching, sculpting, and clustering, whose effects translate into control parameters for genAI pipelines.

Our approach builds on prior work in direct manipulation for AI systems~\cite{masson_directgpt_2024}, extending instrumental interaction frameworks to generative models~\cite{riche_ai-instruments_2025}, treating text as material for co-generation~\cite{shen_texterial_2026}, and developing specialized ideation canvases that move beyond linear chat-based interactions~\cite{marquardt_imaginationvellum_2025}.
We extend these efforts by grounding AI constraint expression in physical material qualities,
aligning with a growing body of HCI research (e.g.,~\cite{rosner_materials_2012, giaccardi_foundations_2015, wiberg_materiality_2013}) that recognizes materiality as a powerful lens for understanding design constraints and possibilities.
A key advantage of this material framing is that materials naturally prompt intuitive gestures, and research demonstrates that gesture-based interactions enable users to encounter, interpret, and sustain abstract meanings through direct, embodied engagement~\cite{tan_embodied_2018,mekler_framework_2019,shen_texterial_2026}.

To demonstrate the feasibility and potential of our approach, we developed \SystemName, a functional direct-manipulation interactive system where users manipulate digital material representations using familiar gestures: pulling and stretching elastic bands, pinching and modeling clay, manipulating attracting and repelling magnet pairs, and piling up and shaping sand.
These interactions then serve as inputs for genAI pipelines (using Stable Diffusion (SD) and ControlNet modules), generating responsive and controllable design outputs. We illustrate \SystemName through technology probes spanning urban design, flower arrangement, and graphic design scenarios. 
Using these probes, we conducted a qualitative study with ten participants of varying design experience to gain insights into both the value and limitations of material metaphor manipulation for AI-assisted design. The study reveals how our approach addresses key challenges in current genAI interfaces. For designers, our approach offers an expressive method that integrates constraints directly into the design process, while for non-designers, the physicality of materials makes design constraints more accessible and inviting.

In summary, our contributions include:
\vspace*{-0.3em}
\begin{enumerate}
    \item an approach mapping materials to design constraints that enables manipulation through familiar material interactions (\autoref{sec:mapping-constraints});
    \item a demonstration of this approach through three technology probes, with design implications for interactive systems that support material metaphor manipulation of design constraints (\autoref{sec:system});
    \item findings from a preliminary user study showing that the effectiveness of material metaphors depends on whether digital representations and gestural interactions match users' expectations and mental models (\autoref{sec:user-study}).
\end{enumerate}
\vspace*{-0.4em}



\section{Related Work}
\SystemName sits at the intersection of design constraints, material 
affordances, and genAI interaction. We review each of these areas in turn, 
situating our contribution within the broader landscape of embodied approaches to 
human-AI co-creation.

\label{sec:background}
\subsection{The Role of Design Constraints}
Design constraints are the boundaries (spatial, structural, relational, or proportional) that define the solution space within which creative work takes shape~\cite{onarheim_opening_2010, stacey_reshaping_2010}.
Constraints play a dual role in creativity, acting as limitations and enablers of creative thought~\cite{stokes_creativity_2005, lawson_how_2006,biskjaer_self-imposed_2013}.

\citet{biskjaer_constraintbased_2014} formalize a constraint-based understanding of design spaces, offering a schema for how designers navigate and manipulate constraints to shape what is possible.
\citet{mose_biskjaer_decisive_2014} show that highly limiting creative decisions can accelerate the design process and propose \textit{decisive constraints} as an analytical concept for understanding creative turning points.

Externalizing and communicating constraints, however, remains a persistent challenge. 
\citet{chevalier_web_2003} show that the ability to articulate and respect constraints varies with expertise and whether constraints are explicitly prescribed, with both novice and professional designers needing better support.
Within HCI, tools like \textit{Scout}~\cite{swearngin_scout_2020} address this by making constraints explicit and manipulable, enabling rapid exploration of design alternatives through high-level constraint specification.
\citet{magnusson_designing_2010} argues that designing constraints is itself a compositional act: defining a system of constraints encapsulates a space for potential expression, establishing the identity and expressive scope of the instrument.
This framing extends naturally to genAI, where the user is similarly composing constraints for the system to follow. 

Yet current genAI interfaces offer limited means for doing so, requiring users to translate rich, multidimensional constraints into linear text prompts~\cite{subramonyam_bridging_2024}. 
This gap between the centrality of constraints in design practice and the lack of support for communicating them to AI systems motivates \SystemName's approach of encoding design constraints through material representations and gestural interaction.

\begin{figure} [t]
\centering
\begin{tikzpicture}
  \node[anchor=south west,inner sep=0] (img)
    {\includegraphics[width=\columnwidth]{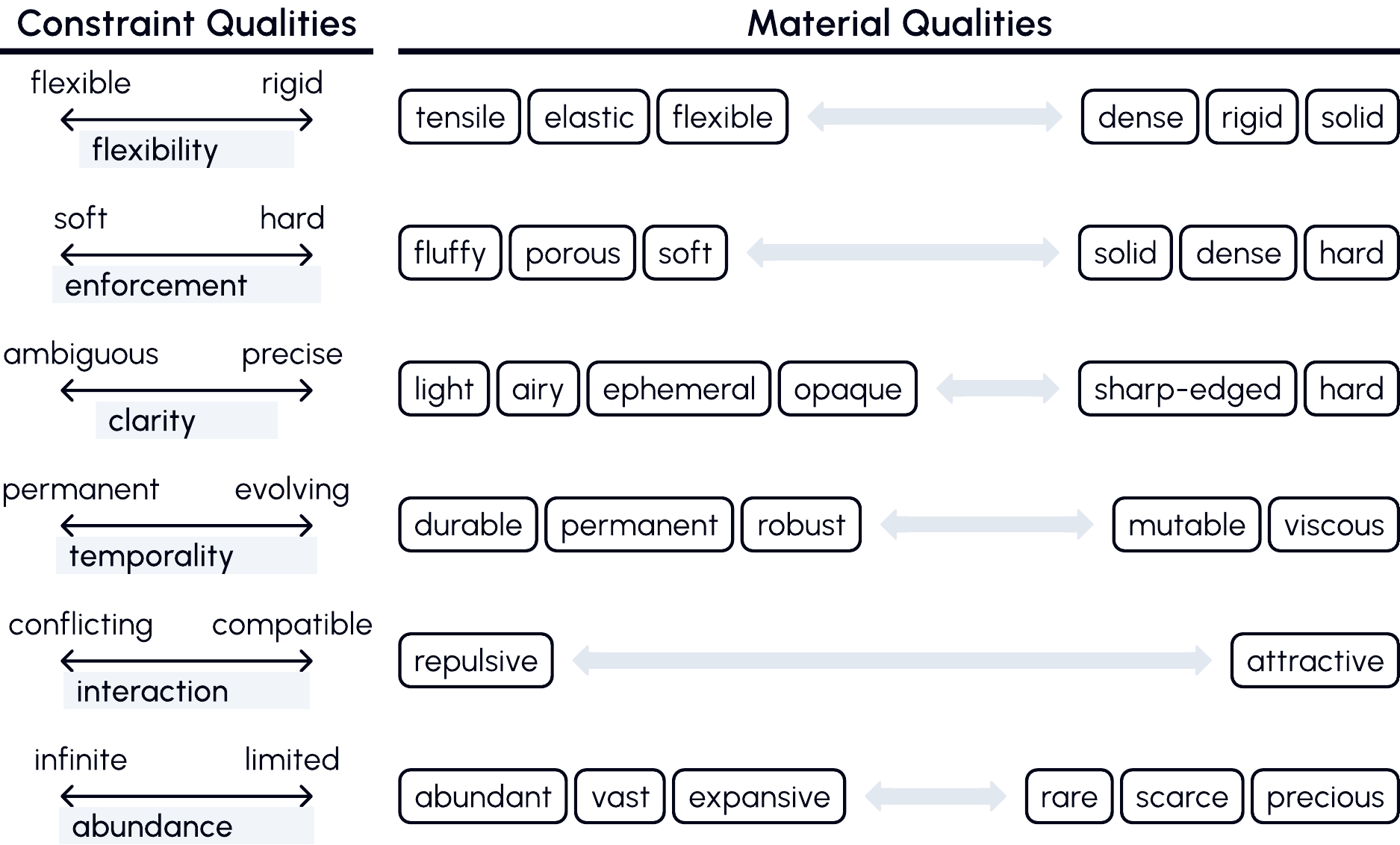}};
  \begin{scope}[x={(img.south east)},y={(img.north west)}]
    \node[font=\tiny] at (0.1844,0.8232) {\cite{elster_ulysses_2000}};
    \node[font=\tiny] at (0.2050,0.6630) {\cite{stacey_reshaping_2010}};
    \node[font=\tiny] at (0.1683,0.5028) {\cite{reitman_heuristic_1964}};
    \node[font=\tiny] at (0.1956,0.3425) {\cite{scudieri_constraint-based_2011}};
    \node[font=\tiny] at (0.1911,0.1823) {\cite{reitman_heuristic_1964}};
    \node[font=\tiny] at (0.1944,0.0221) {\cite{mehta_creating_2016}};
  \end{scope}
\end{tikzpicture}
\caption{Parallel dimensions of constraint qualities (left) and material qualities (right) reveal the cognitive alignment between how constraints and materials are described.}
\Description{
Two parallel columns of bipolar scales. Left, six constraint dimensions: flexibility (flexible to rigid), enforcement (soft to hard), clarity (ambiguous to precise), temporality (permanent to evolving), interaction (conflicting to compatible), and abundance (infinite to limited). Right, clusters of material adjectives aligned to the same rows, such as tensile and elastic opposite dense and rigid, and repulsive opposite attractive. Dotted connectors link each pair.
}
\label{fig:constraint-material-qualities}
\end{figure}

\subsection{Material Affordances \& Gesture Interactions}
HCI scholars increasingly recognize materiality as a lens for understanding the possibilities and constraints of interactive design~\cite{giaccardi_foundations_2015,rosner_materials_2012, wiberg_materiality_2013}.
\citet{giaccardi_foundations_2015} proposed a materials experiences framework, and \citet{karana_material_2015} operationalized this by quantifying materials not only for what they are, but for what they do, express, and make us do.
\citet{dourish_where_2001} provides theoretical grounding through embodied interaction, the idea that tangible computing draws not on everyday artifacts as metaphorical decoration, but on how we experience and act within our environments.
Several systems demonstrate materials as creative digital media, including \textit{SandCanvas}~\cite{kazi_sandcanvas_2011} for multi-touch sand animation and \textit{SandTouch}~\cite{liu_sandtouch_2025} for VR sand art, both showing that material-inspired interactions support rapid, expressive creative work.

Materials are particularly relevant to interaction design because they naturally prompt specific gestures.
\citet{dovencioglu_hand_2022} show that exploratory hand movements vary systematically across materials (i.e., from silk to sand), driven by how materials deform under force, confirming that people possess ingrained gestural knowledge tied to material behaviors.
Gibson's theory of affordances~\cite{gibson_theory_2014} provided the ecological basis for this observation: the action possibilities of an object are perceived directly through its physical properties.
\citet{sun_metaphoraction_2022} formalize how such embodied gestures can carry abstract meanings through a framework for mapping metaphorical meanings to gesture-based interactions.
\citet{renom_exploring_2022} further demonstrate that the technical reasoning people use for physical tool use transfers to digital environments, suggesting that material metaphors allow users to leverage existing physical reasoning rather than learning new interaction vocabularies.

Recent work explores the pairing of materials and gestures as an interaction paradigm for genAI.
\citet{shen_texterial_2026} reimagine text as a material that can be sculpted, grown, and transformed through embodied gestures for LLM-mediated writing. 
This work showed that material metaphors shape users' mental models, foster non-linear workflows, and invite creative exploration beyond text prompting.
We extend this material-metaphor approach to visual designs, mapping everyday materials to design constraints and connecting gestural manipulation to image generation pipelines.

\begin{figure}[t]
    \centering
    \includegraphics[width=\linewidth]{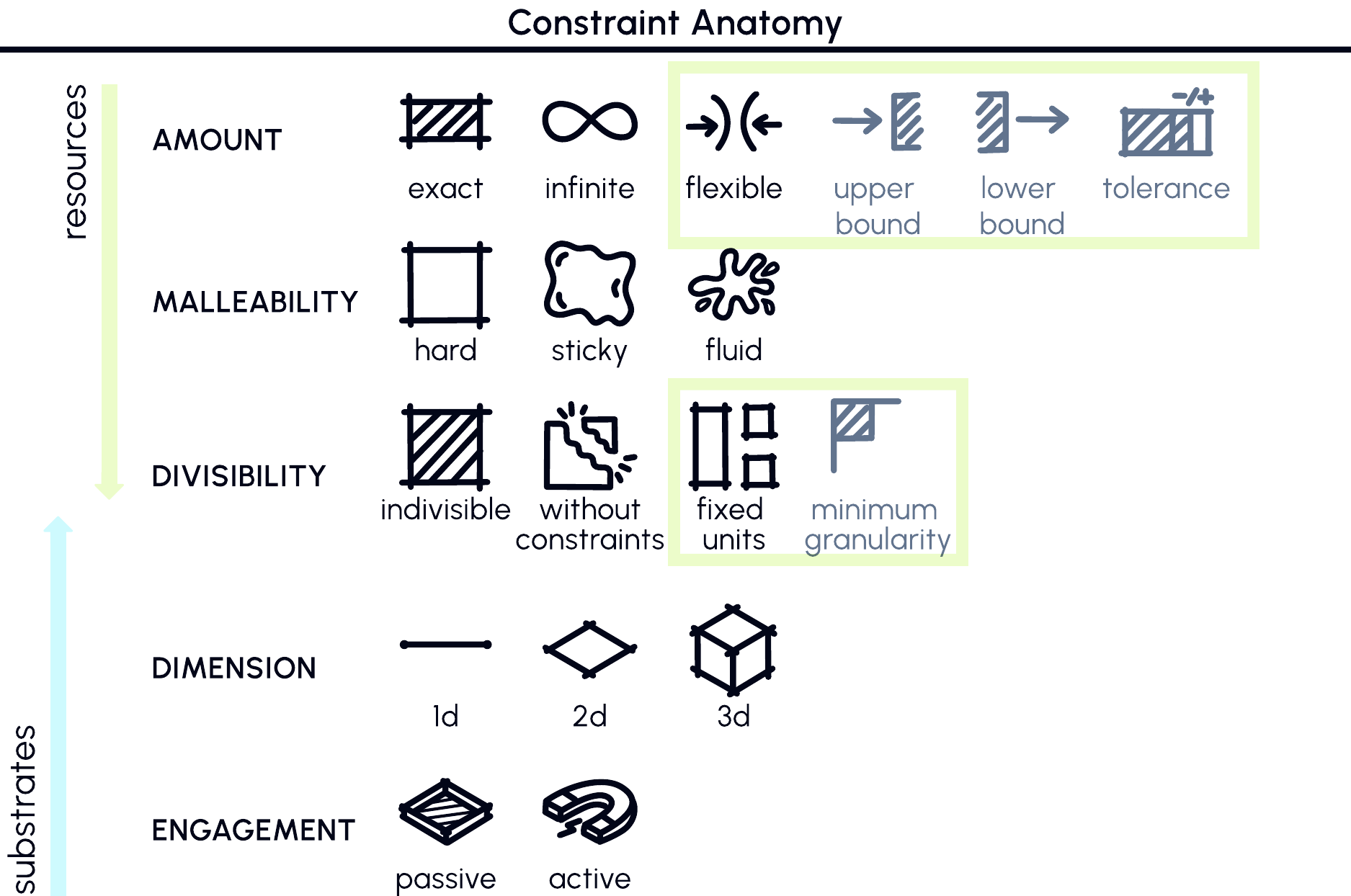}
    \caption{Constraint Anatomy. We conceptualize constraints as the interaction between resources and substrates. 
    Resources are characterized by amount, malleability, and divisibility; substrates by dimensionality and engagement.}
    \Description{
     A chart of the constraint anatomy in two groups. Resources covers amount (exact, infinite, flexible, upper bound, lower bound, tolerance), malleability (hard, sticky, fluid), and divisibility (indivisible, without constraints, fixed units, minimum granularity). Substrates covers dimension (1D, 2D, 3D) and engagement (passive, active). Each value is a small abstract pictogram.
    }
    \label{fig:constraint-anatomy}
    
    \vspace{-1em}
\end{figure}

\subsection{Interacting with GenAI beyond Text}
\label{sec:rw-interacting-with-genai}
Text-to-image prompting has become the dominant paradigm for steering generative models, but its limitations are well documented (e.g.,~\cite{zamfirescu-pereira_why_2023, subramonyam_bridging_2024}).
Users struggle to articulate visual intent~\cite{shi_brickify_2025, liu_beyond_2023, brade_promptify_2023}, iterate through trial-and-error~\cite{liu_design_2022}, and often lack understanding of the models' capabilities~\cite{subramonyam_bridging_2024, mahdavi_goloujeh_is_2024}.
The challenge is particularly acute for design constraints; their multidimensional, interconnected nature makes them difficult to describe and manipulate with text.

\begin{figure*}[t]
    \centering
    \includegraphics[width=\linewidth]{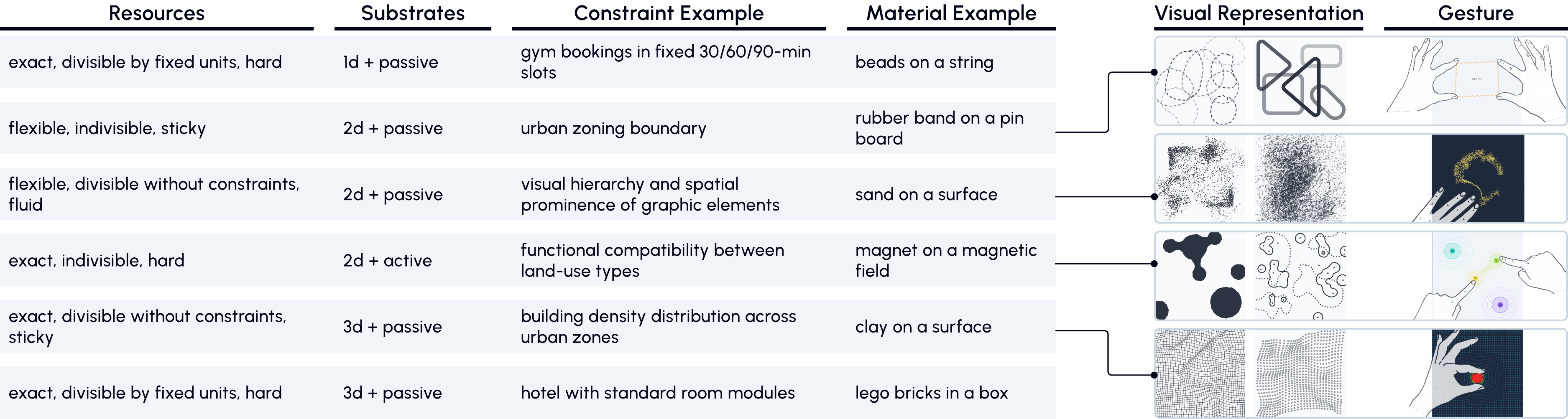}
    \caption{Representative material-to-constraint mappings derived by our framework, showing resource and substrate properties, constraint and material examples, visual representations, and associated gestures.}
    \Description{A six-row table of mappings, with columns for resource properties, substrate properties, a constraint example, a material example, a visual representation, and a gesture. Rows pair gym booking slots with beads on a string, a zoning boundary with a rubber band on a pin board, graphic design hierarchy with sand on a surface, land-use compatibility with a magnet on a magnetic field, building density with clay on a surface, and standard hotel room modules with Lego bricks in a box. Four rows also show a screenshot of the digital material and a line drawing of the hand gesture.}
    \label{fig:mapping-examples}
    
\end{figure*}

A growing body of work addresses these limitations through alternative interaction paradigms.
One approach operationalizes prompt-based interaction as general-purpose tools: \citet{riche_ai-instruments_2025} introduce \textit{AI-Instruments}, embodying prompts as graphical interface objects through reification, reflection, and grounding.
In a complementary direction, \citet{masson_directgpt_2024} map prompts to direct manipulation actions, and \citet{shi_brickify_2025} enable design intent specification through direct manipulation on design tokens.
Other approaches leverage spatial and trajectory-based controls: \textit{TaleBrush}~\cite{chung_talebrush_2022} and \textit{PromptPaint}~\cite{chung_promptpaint_2023} let users manipulate generation trajectories, while \textit{DesignPrompt}~\cite{peng_designprompt_2024} and \textit{WorldSmith}~\cite{dang_worldsmith_2023} use spatial layout as a steering modality.
\citet{marquardt_imaginationvellum_2025} introduced \textit{ImaginationVellum}, a freeform ideation canvas where the canvas itself serves as the prompt, modulating AI generation through spatial proximity, generative strokes, and intent tags.
Together, these systems extend the interaction vocabulary available for steering genAI through spatial arrangement~\cite{marquardt_imaginationvellum_2025}, instrumental interaction~\cite{riche_ai-instruments_2025}, and design tokens~\cite{shi_brickify_2025}.

In fact, direct manipulation has offered non-textual control of design intent long before generative models.
Constraint-based drawing maintains relation limits as the user drags~\cite{gleicher_drawing_1994};
soft selection spreads a local edit outwards with decreasing influence~\cite{igarashi_as-rigid-as-possible_2005};
force fields pull objects together or push them apart~\cite{ahlstrom_modeling_2005};
and particle canvases support granular layout~\cite{kazi_sandcanvas_2011, liu_sandtouch_2025}.
Sketch- and mask-based conditioning has since carried this tradition into diffusion models, including the ControlNet pipelines we build on~\cite{zhang_adding_2023}.

Where most direct manipulation approaches define or extend a set of interaction mechanisms, \SystemName instead leverages the embodied knowledge people already possess about everyday materials (elastic bands, clay, magnets, sand) and the gestures these materials naturally invite.
While prior approaches offer tool-mediated, more precise and granular control, material-elicited gesture interactions are inherently approximate and whole-handed~\cite{dovencioglu_hand_2022, kazi_sandcanvas_2011, liu_sandtouch_2025}.
We argue this quality is not incidental but productive: it lowers the barrier to entry~\cite{bakker_embodied_2012}, supports exploratory engagement~\cite{jacucci_performative_2007}, and affords open-ended creative expression aligned with how people naturally approach early-stage design~\cite{dourish_where_2001}.



\section{Mapping Materials to Design Constraints}
\label{sec:mapping-constraints}

Design constraints are the boundaries---spatial, structural, relational, or proportional---that define the solution space within which creative work takes shape.
Surveying how constraints are discussed across design and cognitive psychology, we observed that researchers characterize them using qualitative language that mirrors how people describe physical materials.
Constraints are spoken of as \textit{flexible} or \textit{rigid}~\cite{elster_ulysses_2000}, \textit{soft} or \textit{hard}~\cite{stacey_reshaping_2010}, \textit{ambiguous} or \textit{precise}~\cite{reitman_heuristic_1964}.
We identify six such dimensions---flexibility~\cite{elster_ulysses_2000}, enforcement~\cite{stacey_reshaping_2010}, clarity~\cite{reitman_heuristic_1964}, temporality~\cite{scudieri_constraint-based_2011}, interaction~\cite{stacey_reshaping_2010,reitman_heuristic_1964}, and abundance~\cite{mehta_creating_2016}---along which both constraints and materials can be described (\autoref{fig:constraint-material-qualities}). 
This overlap suggests a natural cognitive alignment:
a text block that must stay integral shares the same quality language as a hard, rigid material; while decorative graphic elements that can be cropped, stretched, or rearranged to fill empty spaces evoke the fluidity of sand.

This cognitive alignment motivates the use of materials as metaphors for constraints. However, these qualitative descriptions alone are not readily usable for systematically selecting an appropriate material mapping.
Describing a constraint as \say{flexible} does not specify \textit{how} it is flexible, nor does it capture the spatial or interactive context in which the constraint operates.
Inspired by prior work on constraint typologies~\cite{stokes_creativity_2005, mose_biskjaer_decisive_2014, stacey_reshaping_2010}, we propose a complementary lens that conceptualizes constraints as the interaction between \textbf{\textit{resources}}---the elements being constrained---and \textbf{\textit{substrates}}---the space within which constraints operate. 
We characterize resources along dimensions such as amount, divisibility, and malleability, and substrates along dimensions such as dimensionality and engagement (\autoref{fig:constraint-anatomy}).
These dimensions are not intended to be exhaustive; rather they offer a practical vocabulary and an approach for reasoning about the structural properties of constraints, which in turn can inform the selection of materials whose physical behaviors best embody them.
Similarly, the values within each dimension admit further granularity: \textit{flexible}, for instance, can be further decomposed into whether the constraint enforces an upper bound, a lower bound, or permits variation within a tolerance range (\autoref{fig:constraint-anatomy}).
Our goal is not a definitive taxonomy, but a descriptive vocabulary that makes the structural properties of constraints legible and informs material selection.

\begin{figure*}[t]
    \centering
    \includegraphics[width=\linewidth]{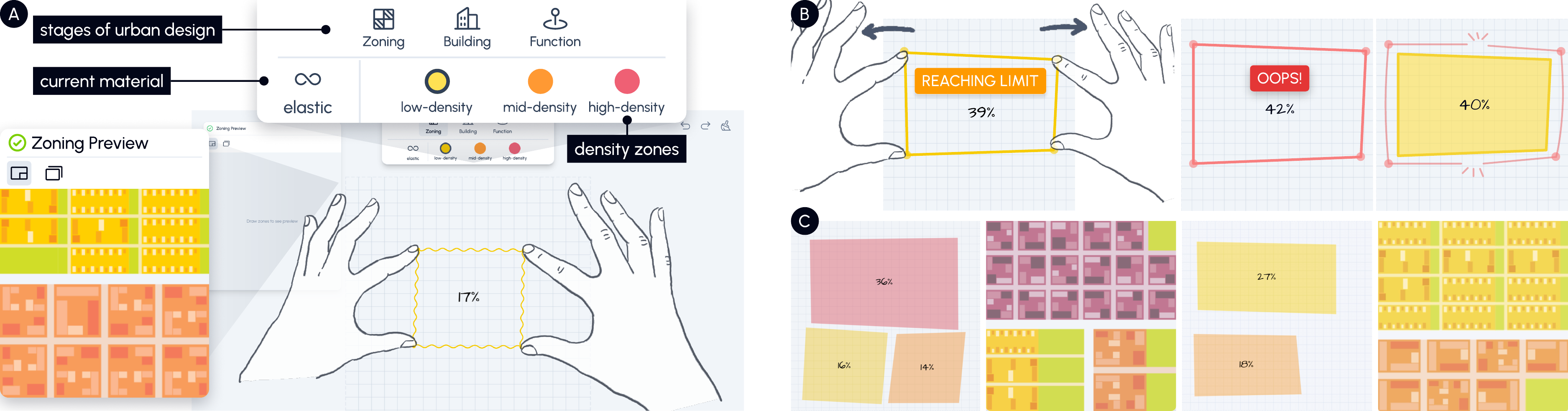}
    \caption{Zoning as elastic bands. 
    (A) Users create zone boundaries via four-finger touch; real-time area percentages are displayed. 
    (B) Bands snap back when stretched beyond allowable limits, providing visual constraint feedback.
    (C) Different zoning configurations produce varied urban layouts, with buildings auto-populated according to density level.}
    \Description{Three panels of the zoning interface. A shows the full screen, with element tabs, density swatches, a zoning preview, and two hands stretching a yellow elastic loop reporting 17 percent coverage. B shows a band at 39 percent labeled "reaching limit," at 42 percent labeled "oops," and snapped back to a permitted 40 percent. C shows six completed zoning configurations, each populated with building typologies matching its density level.}
    \label{fig:elastic}
\end{figure*}

\autoref{fig:mapping-examples} illustrates six representative pairings produced by applying this mapping approach. 
For four of these examples, we show visual representations and a gesture vocabulary: 
rubber bands for flexible zoning boundaries, sand for fluid graphic design hierarchy, magnets for building-use compatibility, and clay for building density 
distribution. 
Together, these examples show how differences across 
constraints translate into materially distinct interactions.
These examples illustrate how the material-inspired lens
generates principled mappings across different constraint types and use cases, and how each material naturally 
elicits gestures that enact the corresponding constraint behavior.



\section{\SystemName: Applying Material Metaphors to Design Constraints}
\label{sec:system}

To explore the conceptual framework presented in \autoref{sec:mapping-constraints}, we developed \SystemName as a proof of concept that applies material metaphors to design constraints through a gestural multi-touch interface.
Our design leverages the inherent affordances of familiar materials---elastic bands, clay, magnets and sand---to elicit natural gestural responses rooted in users' embodied experience. 
Users stretch, mold, push, and scatter digital representations in ways that mirror how they would physically handle these materials, while the system translates these manipulations into meaningful design constraints for a genAI pipeline. 
We present urban design as our primary demonstration scenario, followed by two additional application domains that illustrate the generalizability of the approach.

\subsection{Urban Design: Constraints as Material}
\label{sec:urban}
Urban design involves the planning and organization of physical spaces within cities, balancing multiple competing constraints to create functional, livable environments.
We selected urban design as our primary demonstration scenario because it involves several distinct yet interacting constraint types---such as zoning boundaries, building density, and functional compatibility---that lend themselves well to material metaphor mappings.

Rather than developing a professional urban design tool to replace existing systems, this scenario serves as a technology probe to explore how material affordances can make design constraints more accessible and manipulable through embodied interaction.

\subsubsection{Urban Design Elements and Constraints}
The \SystemName urban design probe focuses on three design elements that exemplify different constraint types commonly found across design domains: flexible boundaries (zoning), resource allocation constraints (building density), and relational constraints (functions).

\noindent\textbf{Zoning} defines permissible land use within designated areas, sometimes expressed as flexible ranges rather than rigid boundaries (e.g., 20--30\% residential allocation for a new development). 
Modern zoning regulations can also be modified over time as development needs change~\cite{elliott_better_2008}.
Such constraints exhibit elasticity: they accommodate variation within bounds but fail when limits are exceeded.

\begin{figure*}[t]
    \centering
    \includegraphics[width=\linewidth]{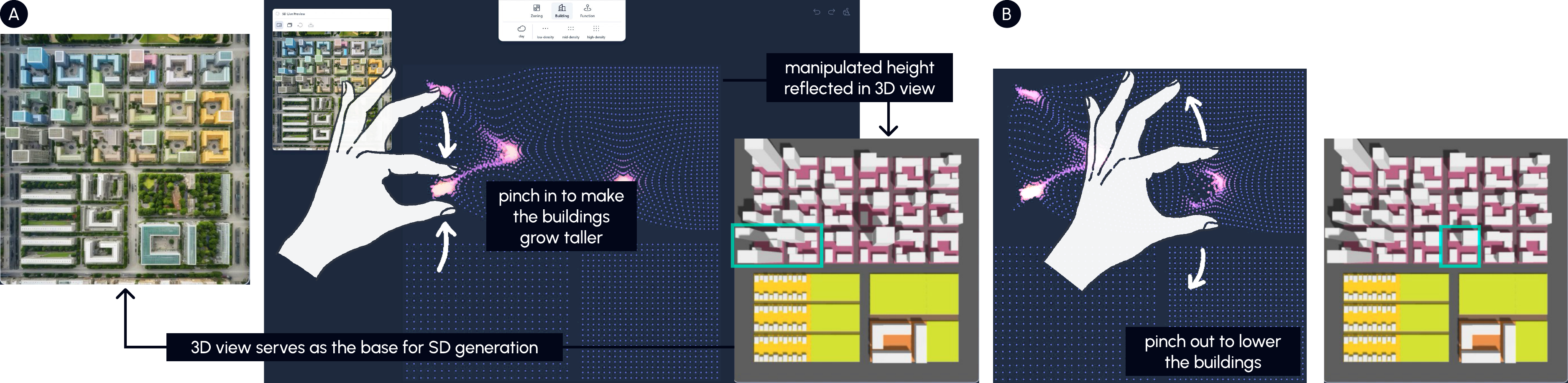}
    \caption{ Buildings as clay. 
    (A) Pinching inward on a dot grid increases building density, reflected in the 3D preview. 
    (B) Pinching outward lowers buildings, redistributing density while preserving total volume. 
    The 3D model serves as input to SD.
    }
    \Description{
     Two panels of density manipulation on a dark canvas of glowing dots. In A, a hand pinches inward to raise building height, with insets showing the change reflected in a 3D urban model and a colored zoning map. In B, the hand pinches outward to lower buildings, with the resulting lower-rise aerial rendering at the right. The dot field brightens where density increases.
    }
    \label{fig:clay}
\end{figure*}

\begin{figure*}[b]
    \centering
    \includegraphics[width=\linewidth]{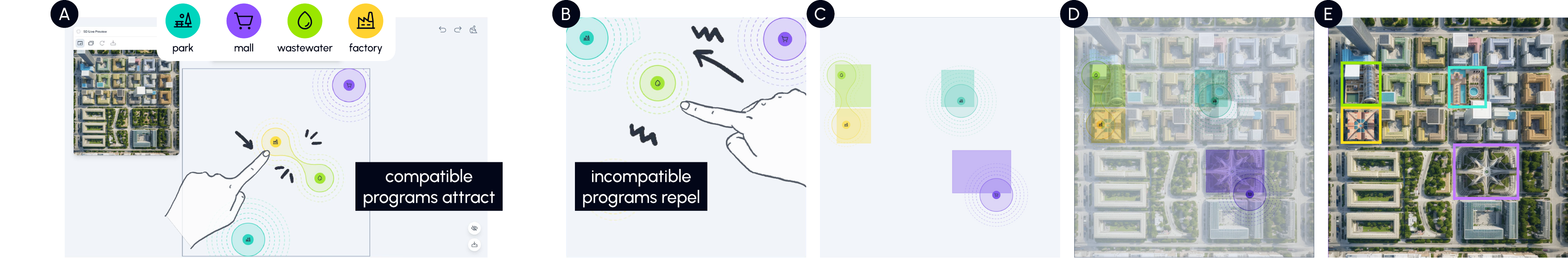}
    \caption{Functions as magnets. (A) Colored circles represent urban functions with attraction/repulsion based on compatibility.
    Compatible functions attract when placed near each other.
    (B) Incompatible programs repel when dragged near each other. (C–E) Selected blocks are inpainted with function-specific imagery to produce the final result.
    }
    \Description{Five panels. A shows four colored magnet icons for park, mall, wastewater, and factory, with a hand dragging one and compatible programs pulling together. B shows incompatible programs pushing apart. C shows the resulting magnet layout with translucent regions marking affected blocks. D and E show the aerial rendering with those blocks inpainted with function-specific imagery, and with colored bounding boxes indicating which block each magnet controls.}
    \label{fig:magnet}
\end{figure*}

\noindent\textbf{Building density} represents the distribution of built volumes, or how concentrated development is within an area~\cite{resch_impact_2016}.
While total volume remains fixed, designers can redistribute density---concentrating height in some zones while keeping lower profiles elsewhere---to create varied urban experiences.

\noindent\textbf{Functions} dictate the spatial relationships between building types and uses.
Some are naturally compatible (e.g., shopping and dining) while others require separation (e.g., industrial and residential), creating attraction and repulsion constraints that shape urban layout. 

\subsubsection{Material Metaphor Mapping}
We map design constraints to material behaviors that exhibit analogous properties:

\noindent\textbf{Elastic Bands for Zoning.} Elastic bands embody flexibility within limits, maintaining form when appropriately tensioned but snapping when overstretched.
This behavior mirrors zoning constraints, which accommodate variation within prescribed bounds but fail beyond acceptable thresholds.
In terms of our constraint anatomy (\autoref{fig:constraint-anatomy}), the zoning area is the \textit{resource}---flexible in amount and indivisible---while the land parcel serves as the 2D passive \textit{substrate}.

\noindent\textbf{Clay for Building Density.} Clay maintains constant mass during shape transformation, embodying urban design's constraint of fixed total building volume that designers can redistribute.
When interacting with clay, users can redistribute the same material volume to create varying density patterns; concentrating clay in one area naturally reduces it elsewhere, reflecting real-world trade-offs.

\noindent\textbf{Magnets for Functions.}
Magnetic attraction and repulsion provide intuitive representation of functional compatibility constraints. 
Compatible urban functions behave like attracting magnet pairs, naturally clustering together, while incompatible ones act like repelling pairs, maintaining appropriate separation distances.
Mapping this to our constraint anatomy (\autoref{fig:constraint-anatomy}), the functions are the \textit{resources}, and the urban layout acts as a 2D active \textit{substrate}, exerting compatible forces that govern where resources can be placed.

\subsubsection{Interface \& Interaction}
The urban design probe interface consists of five main parts (\autoref{fig:elastic}A): a top toolbar with three design element tabs (Zoning, Building, Function), a submenu displaying element-specific interaction options, a central interaction canvas, an SD image preview on the top left, and utility controls on the top right.  
Each design element tab transforms the interface to support corresponding material interactions.

\noindent \textbf{Zoning as Elastic Bands.}
For Zoning elements, users create elastic band loops representing zones through four-finger touch gestures on the canvas (\autoref{fig:elastic}A).
As users move their fingers, the elastic loop updates dynamically, displaying real-time calculations to indicate zoning coverage. 
When users lift their fingers, the system converts the loop into a defined zone boundary.
The elastic bands enforce zoning constraints through material behavior: when a zone exceeds maximum allowable area, the band snaps back to the permitted threshold, providing visual feedback about the design constraint limits (\autoref{fig:elastic}B).
Created zones can be edited via long press or deleted via double-tap (\autoref{fig:elastic}C).
These user-defined zones are automatically populated with pre-generated building typologies based on established urban design practices~\cite{kohn_pedersen_fox_associates_ideal_2014}.
Each density level has contextually appropriate architectural forms: low-density zones include single-family houses and low-rise apartments, mid-density neighborhoods feature mixed-use buildings and townhouses, and high-density regions resemble urban downtowns.
\newline
\newline
\noindent\textbf{Buildings as Clay.}
For Building elements, the canvas transforms into a grid of dots representing malleable clay, where dot density corresponds to the zoning constraints established in the previous tab (\autoref{fig:clay}A).
Low-density zones display sparse dot patterns while high-density zones show tightly packed grids, visually encoding developable volume across areas.
Users manipulate building density through familiar pinching gestures: pinching inward increases building density while pinching outward reduces it (\autoref{fig:clay}A \& B), with changes immediately reflected in the 3D urban model preview (bottom right).
As users interact with the clay, the system sends the updated 3D model to SD, generating photorealistic renderings in the preview panel.

The interface supports two viewing modes: side-by-side (default) with the generated image in the top left corner, and overlay mode where the generated image appears directly on the canvas.
A transparency slider allows users to adjust the opacity of the generated image, enabling examination of relationships between height manipulations and the resulting imagery.
The dot grid employs a sunset color gradient indicating building density.
This creates a heat map effect when overlaid with SD-generated images, helping users understand the correlation between their manipulations and the resulting urban forms.

The interaction technique focuses on manipulating clusters of buildings rather than individual buildings, reflecting urban design's emphasis on neighborhood-scale density patterns. 
The clay metaphor encourages manipulation of the overall building massing distribution across urban blocks, while the material properties implicitly encode the necessary constraints.
\newline
\newline
\noindent\textbf{Urban Functions as Magnets.}
Urban Functions are represented as colored magnets corresponding to four building types: amusement park, shopping mall, wastewater treatment plant, and a textile factory (\autoref{fig:magnet}A). 
The magnetic behaviors are based on functional compatibility---compatible programs (e.g., parks and malls) attract when moved near each other, while incompatible pairs (e.g., parks and factories) actively repel (\autoref{fig:magnet}B \& C).

As users drag magnets across the canvas, the prototype identifies the nearest urban block and updates its functionality visually. 
Target blocks pulse to provide visual feedback during updates, ensuring users understand which areas will be affected. 
Users can toggle overlay mode to view the generated imagery at full canvas size,
with an optional visual aid that highlights urban blocks with corresponding colored boxes, providing precise feedback about which block each magnet will affect.

The four programs are a scoping choice for the probe; future users can easily extend to others such as restaurants, schools, or concert halls by specifying compatibility relationships.

\begin{figure*}[t]
    \centering
    \includegraphics[width=\linewidth]{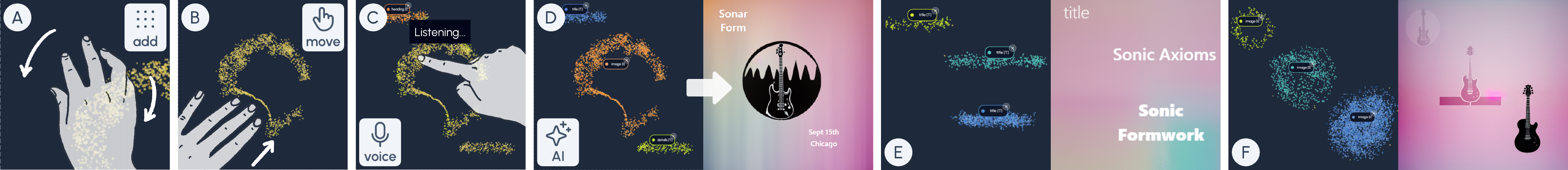}
    \caption{Sand for graphic design. (A) Multi-touch gestures emit sand clusters in ``add'' mode. (B) ``Move'' mode enables reshaping and repositioning. (C) Voice input assigns semantic roles (title, image, logo) with color coding. (D) AI generates the complete poster. (E–F) Cluster size and density control element scale and font weight/opacity.}
    \Description{
    Six panels of the poster workflow. A, a hand emits a cluster of sand particles in "add" mode. B, a hand reshapes a cluster in "move" mode. C, voice input assigns semantic roles and clusters change color by function. D, the tagged clusters beside the generated concert poster. E and F, two further posters showing that larger, denser clusters yield larger and heavier text and more opaque imagery.
    }
    \label{fig:sand}  
\end{figure*}

\begin{figure}[b]
    \centering
    \includegraphics[width=1\linewidth]{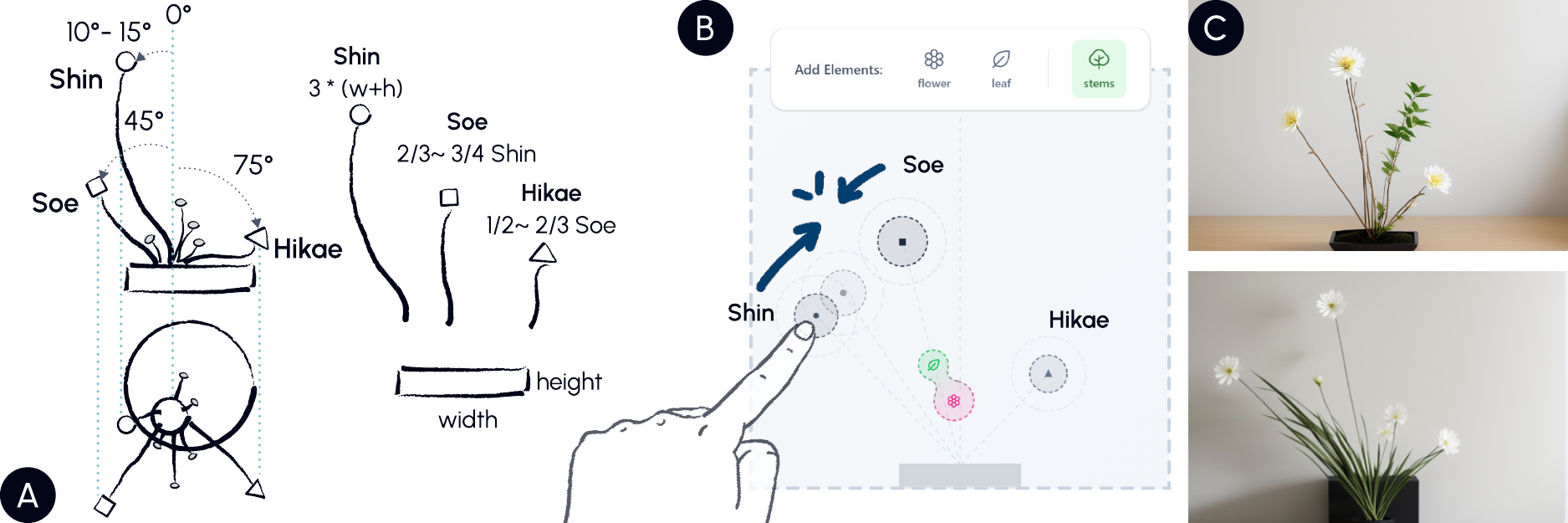}
    \caption{Ikebana as magnets. (A) Traditional angular and proportional rules for Shin, Soe, and Hikae relative to the Utsuwa. (B) The interface encodes these rules in a magnetic system that restricts placement to valid positions. (C) Resulting generated arrangements.}
    \label{fig:ikebana}
    \Description{Three panels. A diagrams the ikebana rules: Shin, Soe, and Hikae radiate from a container at 10 to 15, 45, and 75 degrees from vertical, with proportional lengths annotated relative to the container. B shows the interface, with three labeled magnetic circles above a container rectangle and a hand dragging one. C shows two generated photographs of flower arrangements matching the specified structure.}
\end{figure}

\subsection{Extending to Other Applications}
We illustrate \SystemName's generalizability with two additional scenarios.

\subsubsection{Using Magnets for Ikebana}
Ikebana is the Japanese art of flower arrangement, a practice that treats negative space as equally significant to the floral components, emphasizing that the voids between elements carry as much compositional weight as the materials~\cite{kelly_ikebana_2017}. 
The art form focuses on achieving equilibrium and harmony through careful relationships between organic forms~\cite{kelly_ikebana_2017}. 
Considering these spatial and relational dynamics of ikebana, the magnetic metaphors are one possible material-to-constraint mapping, where attraction represents compatible element relationships and repulsion enforces appropriate spacing. 
\SystemName's magnetic approach allows users to explore ikebana's complex positional rules through intuitive push-pull interactions that mirror the natural tensions and harmonies central to this art form.
\newline
\newline
\noindent\textbf{Design Constraints.}
The three structural elements are known as Shin (heaven), Soe (humanity), and Hikae (earth), each governed by distinct constraints (\autoref{fig:ikebana}A).
Shin is positioned 10--$15^\circ$ from vertical, Soe at $45^\circ$, and Hikae at $75^\circ$~\cite{sato_ikebana_2012}.
Proportionally, Shin's length should equal three times the combined width and height of the Utsuwa (container), Soe should be 2/3--3/4 of Shin's length, and Hikae should be 1/2--2/3 of Soe's length~\cite{sato_ikebana_2012}.
\newline
\newline
\noindent\textbf{Interface \& Interaction.}
The ikebana interface reuses the magnet material, with three magnetic circles for structural elements and a rectangle for the container (\autoref{fig:ikebana}A). 
The magnetic system restricts placement to valid positions, enforcing traditional proportional and angular requirements.

Users can add supplementary elements like leaves and flowers, which exhibit attraction and repulsion behaviors based on composition principles---compatible elements attract while similar types repel to encourage visual variety (\autoref{fig:ikebana}B). 
As the user interacts with the magnetic circles, the arrangement is sent to SD using scribble ControlNet with prompts specific to ikebana aesthetics (\autoref{fig:ikebana}C).

\subsubsection{Using Sand for Graphic Design}
\label{sec:sand-for-graphic-design}
Sand has a long history as an artistic medium, from traditional sand drawing practices to contemporary sand animation performances~\cite{tafler_drawing_2019, kazi_sandcanvas_2011}.
The material has also been appropriated for digital creative applications including \textit{SandCanvas}~\cite{kazi_sandcanvas_2011} and \textit{SandTouch}~\cite{liu_sandtouch_2025}. 
Sand offers an expressive vocabulary of physical gestures that engage the whole hand rather than discrete strokes, creating interactions that are intuitive to learn, quick to execute, and easy to modify---qualities that make sand particularly well-suited for design exploration and iteration~\cite{kazi_sandcanvas_2011}.

Effective graphic design demands careful coordination of focal points, visual hierarchy, and spatial placement, yet communicating these visual relationships through text presents significant challenges, especially when design elements are closely integrated~\cite{shi_brickify_2025}. 
\SystemName's sand metaphor addresses this challenge by allowing designers to directly manipulate spatial layouts through these natural, expressive interactions that support rapid experimentation with compositional arrangements.
\newline
\newline
\noindent\textbf{Design Constraints.}
Graphic design involves spatial and hierarchical constraints that govern element placement and relationships, including visual hierarchy, adequate spacing, and balanced composition weight.
These constraints require simultaneously considering interdependent factors---font sizes must relate to image scales, spacing must accommodate content density, and position must support visual flow while maintaining readability.
Separating spatial layout decisions from content generation allows designers to focus on overall composition before AI systems populate specific elements, enabling rapid exploration of layout variations while maintaining control over spatial relationships.
Following our constraint anatomy (\autoref{fig:constraint-anatomy}), the graphic design elements are the \textit{resources}: flexible in amount, divisible without constraints, and fluid in malleability, while the canvas itself serves as the 2D passive \textit{substrate}.
\newline
\newline
\noindent\textbf{Interface \& Interaction.}
Users begin with a square canvas and can emit sand clusters using multi-touch gestures in the default \say{add} mode (\autoref{fig:sand}A). 
Switching to \say{move} mode allows reshaping and repositioning (\autoref{fig:sand}B). 
Once positioned, users assign semantic roles through voice input—title, main image, logo, description—with each cluster changing color to reflect its function (\autoref{fig:sand}C).
After all clusters have been tagged, users can trigger AI generation to produce the complete graphic design (\autoref{fig:sand}D).

Sand cluster properties directly control generated content through size and density.
Cluster density determines font weight for text elements and opacity for images, while cluster size controls the scale for both text and images (\autoref{fig:sand}E \& F).
Text content is generated via GPT-4 based on design themes and semantic tags (e.g., the \textit{title} of a concert poster), while images are created via SD with transparency support (e.g., the \textit{main image} of a concert poster).

\subsection{Implementation and Pipelines}


\begin{table*}[t]
  \centering
  \small
  \setlength{\tabcolsep}{4pt}
  \setlength{\aboverulesep}{0pt}
  \setlength{\belowrulesep}{0pt}
  \setlength{\extrarowheight}{2.5pt}
  \renewcommand{\arraystretch}{1.15}
    \begin{tabular}{@{}L{0.12\linewidth} M{0.08\linewidth} M{0.22\linewidth} M{0.13\linewidth} M{0.38\linewidth}@{}}
    & \textbf{Material} & \textbf{Constraint} & \textbf{Interaction Parameter} & \textbf{Enforcement Mechanism} \\
    \hline
    \multirow{6}{=}{Constraints enforced in \textit{AI conditioning layer}}
      & \sh Clay    & \sh Building density in urban design           & \sh Building density & \sh Depth map from 3D model $\rightarrow$ ControlNet (Depth) $\rightarrow$ SD \\
      & Magnets     & Functions in urban design                     & Magnet position     & Nearest block's bounding box $\rightarrow$ inpainting mask $\rightarrow$ SD inpainting \\
      & \sh Magnets & \sh Angular and proportional rules in ikebana & \sh Magnet position & \sh Element positions + connecting lines $\rightarrow$ ControlNet (Scribble) $\rightarrow$ SD \\
      & Sand        & Visual hierarchy in graphic design            & Cluster density     & Cluster density $\rightarrow$ image alpha via LayerDiffuse $\rightarrow$ SD \\
    \hline
    \multirow{2}{=}{Constraints enforced in \textit{interaction layer}}
      & \sh Elastic & \sh Zoning in urban design      & \sh Band enclosed area & \sh Band input $\rightarrow$ template-based urban blocks \\
      & Sand              & Element scale in graphic design & Cluster size           & Cluster size $\rightarrow$ post-processing scales generated element \\
  \end{tabular}
  \vspace{1em}
  \caption{Constraints implemented in \textsc{MaterialVellum}, grouped by the layer in which each is enforced.}
  \Description{Table of six material-to-constraint mappings. The first four are enforced in the AI conditioning layer via ControlNet, inpainting, or LayerDiffuse; the last two are enforced in the interaction layer via template selection and post-processing.}
  \label{tab:constraint-enforcement}
  \vspace{-3em}
\end{table*}


\SystemName is implemented as a web-based application with a React.js~\cite{react} frontend and Flask~\cite{flask} backend. 
The frontend uses Matter.js~\cite{matterjs} for physics-based material interactions, GSAP~\cite{gsap} for animations, and Three.js~\cite{threejs} for 3D rendering. 
Image generation runs on a locally hosted SD server (NVIDIA RTX 4090) using multiple ControlNet models (\textit{Depth} and \textit{Scribble})~\cite{zhang_adding_2023}, with model selection depending on the active material metaphor. 
Text generation uses OpenAI's GPT-4 API~\cite{gpt4}. 

We designed technical pipelines for each of the use case applications to leverage multi-touch interactions for manipulation of material representations, which then in turn feed as control parameters into genAI generation (\autoref{tab:constraint-enforcement}).
For example, the pipeline for the urban design use case operates through three sequential stages corresponding to each material interaction.
\textit{Zoning Stage:} User-created elastic band inputs are converted to template-based urban blocks and buildings. 
This 2D urban design plan serves as the foundation for generating a 3D urban model, where building heights are determined by established typologies~\cite{kohn_pedersen_fox_associates_ideal_2014}. 
\textit{Building Stage:} When users manipulate clay to adjust building density, the system captures the updated 3D model and sends it to SD. 
The pipeline first generates a depth map from the image capture of the 3D model; then uses this depth map as ControlNet input to produce photorealistic renderings of the modified cityscape.
\textit{Urban Function Stage:} The generated building imagery serves as the base layer for function-based inpainting.
When users position function magnets, the system calculates bounding boxes for the closest urban blocks, accounting for block dimensions, building density, and camera perspective to ensure accurate coverage of both block dimensions and extruded buildings.
These bounding boxes then form masks that drive targeted inpainting with function-specific prompts (e.g., \say{Drone photography of water amusement park, top down view}). Inpainting results are applied as overlay layers rather than direct edits, allowing blocks to automatically revert to their original appearance when magnets are moved to new locations.



\section{Qualitative User Study}
\label{sec:user-study}
\begin{figure*}[t]
    \centering
    \includegraphics[width=\linewidth]{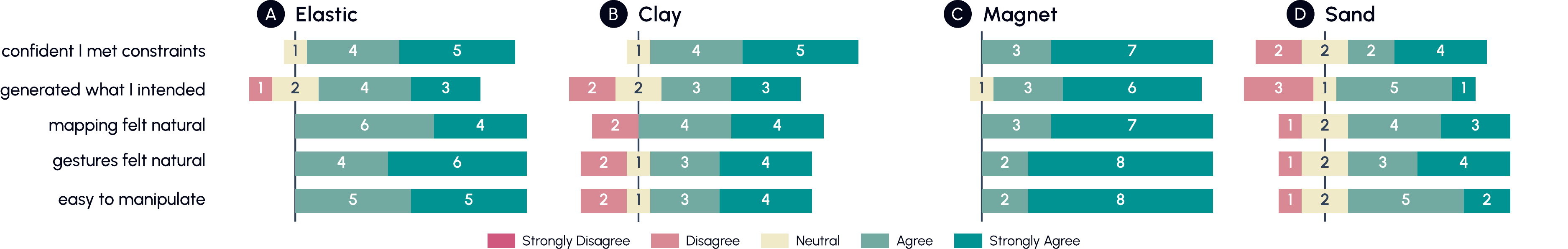}
    \caption{Likert-scale responses (N=10) for each material metaphor across five dimensions, shown top to bottom: constraint adherence, controllability, ease of manipulation, material-to-gesture mapping, and constraint-to-material mapping.}
    \Description{
    Four diverging stacked bar charts, one per material, each with five rows covering mapping, gestures, ease of manipulation, intended output, and constraint confidence across ten participants. Magnet is the most positive, with no negative responses on any dimension and at least seven strong agreements on four of the five. Elastic has no negative responses except a single one on intended output, which also draws two neutrals. Clay carries two negative responses on every dimension except constraint confidence, which is entirely positive. Sand is the most mixed, with negative responses on all five dimensions, including three on intended output.
    }
    \label{fig:likert}
    \vspace{-0.5em}
\end{figure*}

\begin{figure}[b]
    \centering
    \includegraphics[width=\linewidth]{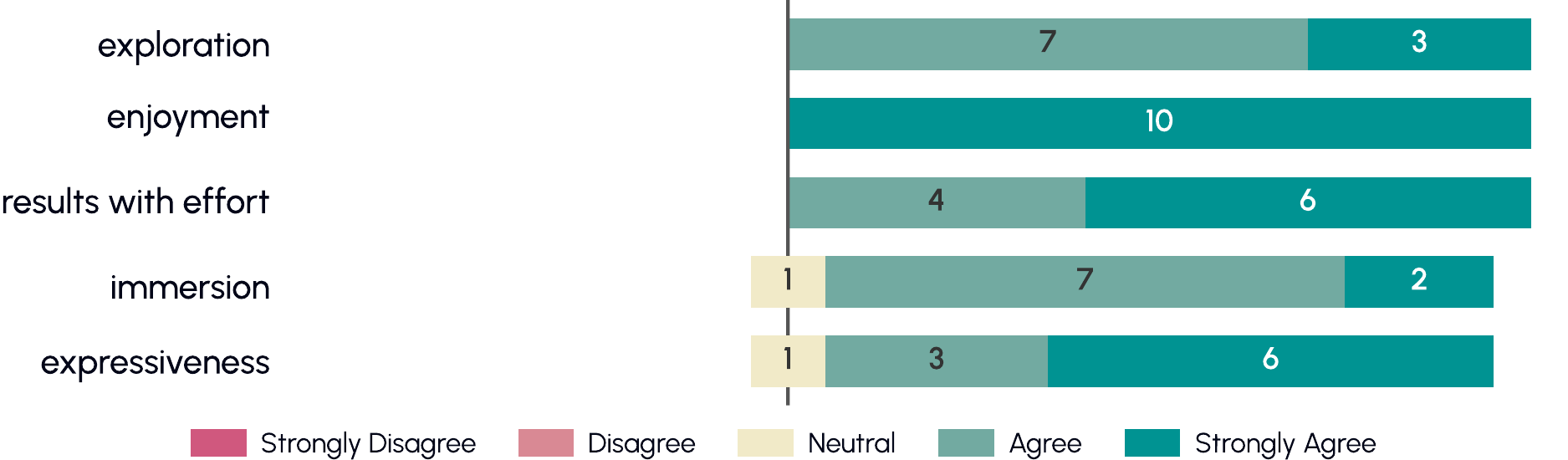}
    \caption{Self-reported Creativity Support Index.}
    \Description{A diverging stacked bar chart of Creativity Support Index responses from ten participants across exploration, enjoyment, results worth the effort, immersion, and expressiveness. Nearly all responses are agree or strongly agree, with one neutral on immersion and one on expressiveness. Enjoyment is unanimous strong agreement.}
    \label{fig:csi}
    \vspace{-0.5em}
\end{figure}

We conducted a qualitative study with ten participants to gather preliminary feedback from both novices and professional designers about the concepts and techniques within \SystemName.

\subsection{Study Design}
\subsubsection{Procedure.}
Sessions took place with participants seated at a Microsoft Surface Studio (28-inch 4.5K display) configured in drafting table mode.
At the beginning of each session, participants signed consent forms agreeing to audio and video recording and completed a brief pre-study questionnaire on demographics and design experience. 
Each session comprised three parts:

\noindent\textbf{Introduction.} The experimenter began by asking participants about their prior experience with image generation tools, then introduced the agenda, and provided a brief overview of \textit{`design constraints'} and \SystemName's material metaphor approach.

\noindent\textbf{Design Activities.} Participants completed two guided design activities using the urban design (\autoref{sec:urban}) and the graphic design probes (\autoref{sec:sand-for-graphic-design}). 
For urban design, participants were asked to imagine themselves in a participatory urban design session, and were required to create three new neighborhood designs: (a) a vibrant downtown with skyscrapers, (b) an elderly-friendly neighborhood, and (c) a neighborhood they would personally want to live in.
For graphic design, participants were asked to create a concert poster using sand.
At the beginning of each activity, the experimenter provided instructions on interaction techniques and UI controls.
The experimenter clarified that there was no evaluation of design output quality, emphasizing that the researcher's goal was to gather genuine reactions and honest feedback about their experiences.
After completing each design activity, the participants were asked to provide immediate reactions to each probe.
    
\noindent\textbf{Feedback.} After completing both design activities, participants filled out a post-study questionnaire with Likert-scale questions assessing their experiences across multiple dimensions.
For each material probe (elastic, clay, magnets, sand), participants rated five dimensions: naturalness of the constraint-to-material mapping, naturalness of the metaphor-prompted gestures, ease of manipulation, whether the system generated what they intended, and their confidence in having adhered to the specified constraints (\autoref{fig:likert}).
The questionnaire also incorporated questions from the Creativity Support Index (CSI) to evaluate \SystemName's overall support for creative activities (\autoref{fig:csi}).
Sessions concluded with the experimenter gathering participants' impressions of the most salient strengths and weaknesses of the experience, perceived value and use case of the approach, suggestions for other design scenarios and materials, and desired improvements.

\subsubsection{Participants.}
Following institutional review board approval, we recruited ten participants (four men, six women) via organizational mailing lists. 
Participants spanned a range of design experience, though the sample skews toward novices (Appendix~\autoref{tab:participant}).
Study sessions were conducted in-person, lasting approximately one hour, with participants receiving \$50 USD gift cards as compensation.

\subsubsection{Data Collection and Analysis.}
Each session was video recorded with audio transcription and screen capture. 
We collected all generated images via \SystemName's built-in logging.
We also collected quantitative data via post-study questionnaires.

Our qualitative analysis focused on identifying high-level themes from the collected data. 
One researcher conducted qualitative open coding~\cite{charmaz_constructing_2006} based on video transcripts and questionnaire responses to develop an initial set of themes, then systematically coded all responses according to these themes. 
To ensure clarity and protect participant privacy, quotes have been slightly modified for concision, grammar, and anonymity.

\subsection{Findings}
Our qualitative analysis, supported by Likert-scale data, revealed six key insights about how participants experienced \SystemName's approach of using material metaphors for forming design constraints to steer generative AI.
\newline
\newline
\textit{\textbf{T1 Embodied material knowledge supports natural gestures.}}
Material metaphors effectively prompted natural gestural interactions that participants found intuitive to learn and remember.
Likert-scale responses on whether the gestures prompted by material metaphors felt natural showed strong agreement for magnets and elastic, with more mixed responses for sand and clay (\autoref{fig:likert}).
After the tutorial, only one participant needed interaction gestures re-explained, suggesting the material affordances effectively communicated their associated gestures.
P6 explained: \sayit{We're so familiar with sand and clay from childhood [\dots] To a point where I feel like [the approach] is quite intuitive.}
The naturalness of the gestures meant that participants could focus on their design goals rather than learning interface mechanics. 
P5 commented: \sayit{For brainstorming, this [gestural interaction] feels much more natural than text prompting.}

Not all material mappings felt equally natural.
P6 identified a disconnect between the pinch gesture and their embodied understanding of sand and clay: \sayit{In the real world, you don't pinch clay or sand. You use your whole hand to morph it sideways. [Sand and clay] should be more of a whole-hand interaction.}


\begin{table*}[t]
  \centering
  \small
  \setlength{\tabcolsep}{4pt}
  \setlength{\aboverulesep}{0pt}
  \setlength{\belowrulesep}{0pt}
  \setlength{\extrarowheight}{2.5pt}
  \renewcommand{\arraystretch}{1.15}
  \begin{tabular}{@{}M{0.17\linewidth} M{0.23\linewidth} M{0.54\linewidth}@{}}
    \textbf{Application Domain} & \textbf{Suggested Materials} & \textbf{Participant Examples} \\
    \toprule
    \sh Interior Design & \sh Elastic band for layout & \sh \sayit{When people are moving and need to rearrange their furniture.} (P6) \\
    Fashion Design & Clay, elastic for garment flow & \sayit{Specify the flow of fabric for shirts or pants.} (P2) \\
    \sh Graphic Design & \sh Magnets for color palettes & \sh \sayit{It would be great if [the magnets] can tell me I'm designing something wrong, like putting colors together that are colorblind-unfriendly or culturally insensitive.} (P6) \\
    Game Design & Magnets for character and maps & \sayit{Character optimization [\dots] making fantasy maps for tabletop games.} (P5) \\
    \sh Scientific Visualization & \sh Sand for layout & \sh \sayit{Figure design for research, faster than Figma for iterations.} (P8) \\
    Nutrition & Magnets for nutrient interactions & \sayit{Iron and calcium would repel since they affect absorption.} (P5) \\
    \sh Scheduling & \sh Magnets for time conflicts & \sh \sayit{Determine what things go together or conflict with each other.} (P5) \\
    Traffic Separation & Magnets for pedestrian safety & \sayit{You can also represent how close pedestrians are to cars.} (P7) \\
  \end{tabular}
  \vspace{1em}
  \caption{Suggested materials for different application domains, with participant examples.}
  \Description{Table of eight application domains, each paired with a suggested material metaphor and a supporting participant quote.}
  \label{tab:suggested-material}
  \vspace{-3em}
\end{table*}


\textit{\textbf{T2 Material metaphor effectiveness varies across participants and contexts.}}
Participants demonstrated distinct preferences across the four material metaphors, with both Likert-scale responses (\autoref{fig:likert}) and qualitative feedback revealing differences in how each metaphor was experienced.
While participants like P1 appreciated the overall approach, \sayit{I like the metaphors a lot, it's a proper way to illustrate the relationship between physical engagement and design,} individual mappings resonated differently with different users.

\textit{Magnets} received the most uniformly positive ratings (\autoref{fig:likert}).
P9 explained: \sayit{My favorite feature was the magnets. I really like the concept of different city areas going well together or not [\dots] it's obvious and effective.}
P6 added: \sayit{I enjoyed the magnet interaction most because I could clearly map my changes to image changes.}
P2 found them engaging for creating realistic environments: \sayit{The magnets are the most fun because it feels like designing a real world environment.}

\textit{Elastic} was rated positively on most dimensions, with participants appreciating the conceptual mapping but expressing some frustration with implementation limitations.
P2 captured this duality: \sayit{I thought the rubber band was really fun [\dots]  but frustrating because it mapped to a template rather than my exact drawings.}

\textit{Clay} and \textit{Sand} showed more polarized responses.
P8 said: \sayit{Clay was what I enjoyed most because the translation from gesture to results has the clearest feedback.}
While P9 felt \sayit{the clay visualization had no correlation to actual clay.}
For \textit{Sand}, P10: \sayit{I like the sand concept, I've never thought of using sand this way, but it reminds me of how kids draw in sand, which feels intuitive.}
P8: \sayit{It's satisfying to put sand onto the canvas. It gives me control and lets me insert my opinion.}
P7 said \sayit{I like this abstract approach to poster design because it helps me think about visual hierarchy and prominence.}
However, P7 also mentioned sand's limits: \sayit{Sand is too abstract for controlling text, which has many visual parameters: boldness, font size, line spacing.}

P4 articulated the underlying tension between clarity and metaphorical richness \sayit{This is a paradoxical design challenge: if you make everything crystal clear with one-to-one mapping, you lose the metaphor. If it's too vague, people struggle with the mental model. The beauty lies somewhere in between [\dots] you give people some guidance while leaving room for discovery.}
\newline
\newline
\textit{\textbf{T3 Controllability for generative AI differs across design contexts.}}
Participants' satisfaction with AI generation outputs varied considerably across materials and design contexts (\autoref{fig:likert}), highlighting tensions between constraint-based design and creative flexibility.
Some participants found constraints helpful for making decisions without domain expertise.
P6: \sayit{Urban design is something I have no knowledge of, but magnets helped me be less aware of my lack of knowledge yet still design correctly.}
P6 also valued the balance: \sayit{This system hits a nice middle ground where I can specify that I want a particular area to be a park, but leave everything else up to the AI's imagination.}
However, others wanted more flexibility.
P2 noted: \sayit{I tried to draw specific designs but got templates that were only somewhat similar.}
P5 echoed this desire for greater precision: \sayit{The ability to interact directly is really nice, though I find myself wanting finer control.}
\newline
\newline
\textit{\textbf{T4 Novice-friendly approach that supports early-stage ideation and educational purposes. }}
\SystemName proved particularly effective for novice users and educational contexts.
P7: \sayit{The tool has educational value for learning design constraints through interaction. For example, I learned that parks and shopping malls complement each other.}
The staged design process was especially beneficial for building novice confidence. 
P7: \sayit{I appreciate the three-step guided process and how each step builds on the previous one. This naturally leads me to consider how density relates to building heights and their corresponding functions.}
P7 further noted how the constraints reduced cognitive load: \sayit{The rubber band gives me confidence that I could focus on my design goals while the system handles the constraints.}

Participants repeatedly identified educational applications, particularly for children and community engagement.
P4: \sayit{This is perfect for community engagement with non-technical participants who have ideas but lack technical skills. They can experiment without needing precision.}
P7 envisioned specific educational contexts: \sayit{This tool is ideal for early-stage conceptualization—understanding the general vibe and what the design looks like from a distance. It's particularly good for non-professionals and novices who don't need to think about constraints intentionally. I could see this being used educationally, even with kids in museums.}
\newline
\newline
\textit{\textbf{T5 Playful engagement through embodied interaction. }}
The hands-on nature of the interface resulted in high engagement levels. 
All ten participants strongly agreed with the statement \sayit{I was very engaged/absorbed in the activity---I enjoyed it and would do it again}. 
P5: \sayit{Being able to use your hands to make things is innately satisfying [\dots] I want to keep playing with it.}
P6: \sayit{It's a bit more physically demanding than using a mouse [\dots] but the embodied nature and the fact that you're having a ton of fun makes the extra effort worth it.}
P3 appreciated the sensory dimension: \sayit{It provides visual sensory stimulation that's soothing.}
P5, a former landscape architect, contrasted \SystemName with professional tools: \sayit{I really like this kind of physical, embodied interaction. It's much more fun than using AutoCAD.}
\newline
\newline
\textit{\textbf{T6 Example applications and potential beyond demonstrated scenarios.}}
Participants readily identified application domains beyond urban design and graphic design (\autoref{tab:suggested-material}).
The breadth and specificity of these suggestions indicate that the material framing transferred conceptually: participants could describe constraints in other domains in material terms and imagine gestures for manipulating them.
As P6 expressed: \sayit{This notion of embodied interactions could work for just about any genAI scenario, because how I use genAI today is still mostly text-based, and it's very limiting. Creativity is always bigger than whatever text can describe.}



\section{Discussion}
\label{sec:discussion}

\textit{\textbf{Material Metaphor Effectiveness and Its Limits.}}
Our findings reveal that no single material metaphor is universally effective.
Magnets received consistently high ratings, likely because the attract/repel logic maps cleanly onto compatibility relationships with minimal interpretive ambiguity. 
Clay and sand, by contrast, elicited polarized responses.
Our findings suggest that the effectiveness of a material metaphor depends not only on the structural fit between material affordances and constraint type, but also on individual users' embodied experiences with those materials.

A core design tension underlies these differences: the more literal the mapping between physical material and digital behavior, the easier users can form a mental model, but overly faithful mappings risk introducing false equivalences. 
Physical clay, for example, cannot be unsculpted, yet reversibility is expected and necessary in a digital design tool.
As P4 reflected, a perfect one-to-one correspondence eliminates the metaphor's generative power, while excessive vagueness leaves users without a workable mental model.
This reflects what~\citet{smith_experiences_1986} called the \sayit{tension 
between literalism and magic}---the design challenge of deciding 
when the metaphorical vehicle should give way to capabilities 
of the digital system~\cite{dourish_where_2001}.
Our results suggest that the most productive material metaphors are those that preserve enough physical fidelity to be immediately graspable while deliberately departing from literal behavior where the design task demands it.
\newline
\newline
\textit{\textbf{Material Metaphors Complement Rather Than Replace Other Interaction Paradigms.}}
Several participants expressed a desire to combine material interactions with other modalities (drawing, speech, and text input), particularly when they wanted finer-grained control over generated outputs.
Participants observed that material metaphors are best suited for the early, exploratory phases of a design process, where approximate, whole-handed gestures align well with the goal of rapidly sketching out constraint relationships and spatial arrangements.
This positioning resonates with Texterial \cite{shen_texterial_2026}, which similarly found that material-based interactions complement rather than replace conventional writing interfaces. 
In our case, material metaphors excel at externalizing high-level constraint structures (e.g., which zones attract or repel, how density distributes across a site) but are inherently imprecise for specifying exact values.
The same holds against the direct manipulation techniques discussed in \autoref{sec:rw-interacting-with-genai}. 
Soft selection, force fields, and mask-based conditioning are conventionally exposed through explicit parameters that users can set, whereas we deliberately trade that precision for approximate, whole-handed gestures.
Future systems could therefore position material interactions as one layer within a multi-modal toolkit, allowing users to transition fluidly from gestural exploration to text- or drawing-based refinement as their design intent advances.
\newline
\newline
\textit{\textbf{Technical Challenges and Open Questions.}}
Translating gestural input into genAI control parameters remains highly scenario-specific in the current system. 
Each material-constraint pairing required manual mappings from gesture input to system prompts and ControlNet parameters.
Systematizing this translation through a declarative grammar that maps gesture input to prompt modifiers is a promising future work direction.
The reverse direction, propagating model state back into material behavior, remains unexplored: clay could stiffen where sampling variance is high, or sand density shift with a diffusion model's cross-attention maps. Whether such feedback clarifies model state or undermines the metaphor's predictability is an open question.
\newline
\newline
\textit{\textbf{Resources, Substrates, and Material Decomposition.}}
Our framework conceptualizes constraints as interaction between resources and substrates; this decomposition then guides material selection.
This lens applies well to spatial and proportional constraints, where resources are tangible---but constraints with abstract resources (temporality, logical exclusions) lack obvious physical substrates. 
Yet, the resource-substrate vocabulary is not inherently spatial, offers structured ways to reason about constraint anatomy, inviting experimentation with non-obvious mappings leveraging materials for abstract constraints.

Our constraint anatomy admits substrates of one to three dimensions (\autoref{fig:constraint-anatomy}), but every substrate in our probes is two-dimensional.
Building height is volumetric, yet users express it by pinching a planar grid whose depth is conveyed through color.
Richer surfaces could align a material's dimensionality with its constraints; however, manipulating 3D content through 2D input is a long-standing problem~\cite{hancock_shallowdepth_2007} and depth carries its own perceptual costs~\cite{laviolajr_3d_2017}.
\newline
\newline
\textit{\textbf{Study Design Limitations.}}
Our themes were developed by one researcher and reflect one interpretive perspective.
Single-session studies of a novel tactile interface are subject to novelty~\cite{koch_novelty_2018} and demand~\cite{dell_yours_2012} effects, so our engagement ratings should be read as initial reactions rather than durable evidence.
We also did not include a baseline condition: \SystemName is a technology probe intended to surface reactions to a design space, and we make no claim that material metaphors outperform text prompting or established direct manipulation techniques.
Empirical comparison against these techniques is an important next step, and should evaluate not only task performance but also qualities like enjoyability and memorability, which are where we expect the value of a familiar material metaphor to lie.



\section{Conclusion}
\SystemName demonstrates how physical material affordances can function as an expressive lens for articulating design constraints when working with genAI. 
Our framework offers an initial approximation of how material properties may be mapped to constraint qualities through resources and substrates.
As genAI becomes increasingly central in creative practice, material metaphors suggest one path toward interactions grounded in embodied human experience.
Looking forward, we hope this work inspires exploration of alternative material-to-AI interaction mappings, advancing more creative material-grounded forms of human-AI co-creation.



\begin{acks}
We thank the participants of our user study for taking the time to provide feedback on \SystemName as well as the reviewers of this submission for concrete suggestions to improve this paper.
We also thank Jocelyn Shen, Rima Cao, Ryan Yen, Jude Rayan, Arnavi Chheda-Kothary, Ann Paradiso, Andrew D. Wilson and Jen Fox for their feedback throughout this project.

\end{acks}


\newpage
\bibliographystyle{ACM-Reference-Format}




\appendix




\begin{table*}[h]
  \centering
  \small
  \setlength{\tabcolsep}{5pt}
  \setlength{\aboverulesep}{0pt}
  \setlength{\belowrulesep}{0pt}
  \setlength{\extrarowheight}{2.5pt}
  \renewcommand{\arraystretch}{1.15}
  \rowcolors{2}{white}{rowgray}
  \begin{tabular}{@{}llllll@{}}
    \textbf{PID} & \textbf{Age Range} & \textbf{Gender} & \textbf{Occupation} & \textbf{Design Experience} & \textbf{Image Generation Tools} \\
    \toprule
    P1  & 25 to 34 & Woman & Software Engineer (prev. Architect, 6+ years)           & Intermediate         & DALL-E, Midjourney \\
    P2  & 25 to 34 & Man   & Software Engineer                                  & Intermediate         & DALL-E, Stable Diffusion \\
    P3  & 35 to 44 & Woman & Executive Assistant                                & No design experience & None \\
    P4  & 25 to 34 & Woman & PhD Student                                        & Basic/Beginner       & None \\
    P5  & 25 to 34 & Man   & Researcher (prev. Landscape Architect, 6+ years)        & Intermediate         & Midjourney, Adobe Firefly, Stable Diffusion \\
    P6  & 25 to 34 & Man   & PhD Student                                        & No design experience & DALL-E, Stable Diffusion \\
    P7  & 25 to 34 & Woman & PhD Student                                        & Basic/Beginner       & Stable Diffusion, DALL-E \\
    P8  & 18 to 24 & Woman & PhD Student                                        & Basic/Beginner       & DALL-E, Midjourney, Stable Diffusion \\
    P9  & 55 to 64 & Man   & Technical Program Manager                          & No design experience & Bing, Copilot \\
    P10 & 18 to 24 & Woman & Student                                            & Basic/Beginner       & DALL-E, Midjourney, Stable Diffusion \\
  \end{tabular}
  \caption{Participant demographics, design experience, and prior use of image generation tools.}
  \Description{Table of ten study participants listing age range, gender, occupation, self-reported design experience, and image generation tools used previously.}
  \label{tab:participant}
\end{table*}



\begin{figure*}[t]
    \centering
    \includegraphics[width=\linewidth]{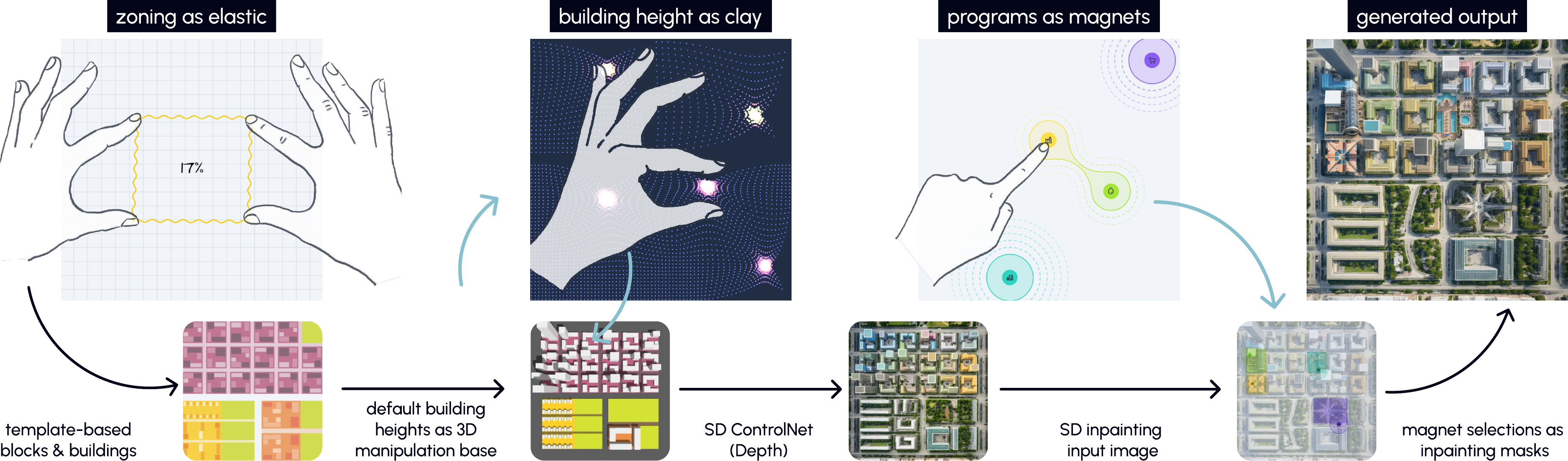}
    \caption{Technical pipeline for the urban design probe. Zoning inputs generate template-based 3D models; clay-based height manipulation feeds depth maps to ControlNet for photorealistic rendering; magnet-placed functions create inpainting masks for targeted image generation.
    }
    \Description{A two-row pipeline diagram for the urban design probe. The top row shows four interface states: hands stretching an elastic zoning loop, a hand pinching a clay dot grid, a hand dragging magnets, and the final aerial rendering. The bottom row shows the corresponding stages: zoning input becomes template-based blocks and a 3D manipulation base, clay manipulation produces a depth map for Stable Diffusion with Depth ControlNet, and magnet selections become inpainting masks for the final output.}
    \label{fig:pipeline}
\end{figure*}


\end{document}